\documentstyle[psfig]{aipproc2}

\begin{document}
\title{CMB and Molecules at High Redshift} 

\author{F. Combes}
\address{Observatoire de Paris, DEMIRM, \\
61 Av. de l'Observatoire, F-75 014 Paris, France}


\maketitle

\begin{abstract}
It becomes possible now to detect cold molecules at high redshift in the
millimeter domain. Since the first discovery in 1992 by Brown and van den Bout 
of CO lines at $z=2.28$ in a gravitationally lensed starburst galaxy,
nearly ten objects are now known to possess large quantities of molecular gas
beyond $z=1$ and up to $z \sim 5$, through millimeter and sub-millimeter 
emission lines. Even more objects have been detected in their
 continuum dust emission, and a few galaxies through millimeter absorption 
lines in front of quasars at $z \le 1$.

The continuum dust emission is the most easily detected: for a starburst
dust at T$_d \sim$ 60 K, the emission peaks around 60 $\mu$m, and falls
as $\lambda^{-4}$ at longer wavelengths. In the mm domain,
the emission is then stronger for the more redshifted objects.
 For the CO lines, the situation is less favorable, and the reported
detections are helped by gravitational amplification. 
The increase of the CMB temperature T$_{bg}$ with redshift helps
the rotational line excitation (especially at high $z$), 
but not its detection.

Absorption in front of quasars is a more sensitive probe 
of cold gas at high redshift, able to detect individual clouds
of a few solar masses (instead of 10$^{10}$ M$_\odot$ for emission).
Taking advantage of the small size of the QSO, very high spatial
resolution (of the order of milli-arcsec) can be achieved, and
high spectral (30m/s) resolution, due to the heterodyne technique.
The sampled column-densities range 
between N(H$_2$)= 10$^{20}$ et 10$^{24}$ cm$^{-2}$.
The high sensitivity allows to detect a multitude of 
molecular lines in a single object (HCO$^+$, HNC, HCN, N$_2$H$^+$, 
C$^{18}$O, CS, H$_2$CO, CN, CCH, H$_2$S etc....), and compare the 
chemistry with the local one, at $z$ = 0.
From the diffuse components, one can measure the 
cosmic black body temperature as a function of redshift. The high
column densities component allow to observe important molecules
not observable from the ground, like O$_2$, H$_2$O and LiH for example.

All these preliminary studies carry a great hope for what will be observed
with future millimeter instruments, and some perspectives are given.
\end{abstract}

\section{Millimeter CO emission lines at high redshift}

This is a rapidly evolving domain, and at present, only 8 systems are
published (cf Table 1). The search of CO lines at
high $z$ has been triggered by the detection of the CO(3-2) line in 
emission in the Faint IRAS source
F10214+4724 at $z = 2.28$  by Brown \& Vanden Bout (1991, 1992).
At this time, it was a redshift 30 times larger than that of the most
distant CO emission discovered in a galaxy.  The H$_2$ mass derived was
reaching 10$^{13} h^{-2}$ M$_\odot$, with the standard CO-H$_2$ conversion
ratio, a huge mass although the FIR to CO luminosities was still
compatible with that of other more nearby starbursts. Since then, the 
derived H$_2$ mass has been reduced by large factors, both with better
data and realizing that the source is amplified through a gravitational 
lens by a large factor (Solomon et al 1992, 1997).

After the first discovery, many searches for other candidates took place,
but they were harder than expected, and only a few,
often gravitationally amplified,
objects have been detected: the lensed Cloverleaf quasar
H 1413+117 at $z=2.558$ (Barvainis et al. 1994),
the lensed radiogalaxy MG0414+0534 at $z=2.639$ (Barvainis et al. 1998),
the possibly magnified object
BR1202-0725 at $z=4.69$ (Ohta et al. 1996, Omont et al. 1996),
the amplified submillimeter-selected hyperluminous galaxy SMM02399-0136
(Frayer et al. 1998), at $z=2.808$, and the magnified BAL quasar APM08279+5255,
at $z=3.911$, where the gas temperature derived from the CO lines is
$\sim$ 200K, maybe excited by the quasar (Downes et al. 1998).
Recently Scoville et al. (1997) reported the detection of the first
non-lensed object at $z=2.394$, the weak radio galaxy 53W002,
and Guilloteau et al. (1997) the radio-quiet quasar BRI 1335-0417, at $z=4.407$,
which has no direct indication of lensing.
If the non-amplification is confirmed, these objects
 would contain the largest molecular contents known
(8-10 10$^{10}$ M$_\odot$ with a standard CO/H$_2$
conversion ratio, and even more
if the metallicity is low).
The derived molecular masses are so high that H$_2$ would constitute
between 30 to 80\% of the total dynamical mass (according to the unknown
inclination), if the standard CO/H$_2$ conversion ratio was adopted.
The application of this conversion ratio is however doubtful, and it is
possible that the involved H$_2$ masses are 3-4 times lower (Solomon
et al. 1997). 

\begin{table}[h]
\caption[ ]{CO data for high redshift objects}
\begin{flushleft}
\begin{tabular}{lllclcl}  
Source    & $z$   &  CO  & S  & $\Delta$V& MH$_2$   & Ref  \\
          &       &line  & mJy  & km/s & 10$^{10}$ M$_\odot$    &           \\
\hline
F10214+4724 & 2.285 & 3-2  & 18 & 230  & 2$^*$      &  1   \\
53W002      & 2.394 & 3-2  &  3 & 540  & 7      &  2   \\
H 1413+117  & 2.558 & 3-2  & 23 & 330  & 6      &  3   \\
MG 0414+0534& 2.639 & 3-2  &  4 & 580  & 5$^*$      &  4   \\
SMM 02399-0136&2.808& 3-2  &  4 & 710  & 8$^*$      &  5   \\
APM 08279+5255&3.911& 4-3  &  6 & 400  & 0.3$^*$    &  6   \\
BR 1335-0414& 4.407 & 5-4  &  7 & 420  & 10         &  7   \\
BR 1202-0725& 4.690 & 5-4  &  8 & 320  & 10         &  8   \\
\hline
\end{tabular}
\end{flushleft}
$^*$ corrected for magnification, when estimated\\
Masses have been rescaled to $H_0$ = 75km/s/Mpc. When multiple images
are resolved, the flux corresponds to their sum\\
(1) Solomon et al. (1992), Downes et al (1995); (2) Scoville et al. (1997);
(3) Barvainis et al (1994); (4) Barvainis et al. (1998); (5) Frayer et al.
(1998); (6) Downes et al. (1998); (7) Guilloteau et al. (1997);
(8) Omont et al. (1996)
\end{table}

It is surprising that very few starburst galaxies have been detected in
the CO lines at intermediate redshifts (between 0.3 and 2), although many have
been observed (e.g Yun \& Scoville 1998, Lo et al 1999). A possible explanation
is the lower probability of magnification by lenses in this range
(cf Figure 1).

\begin{figure}
\centering
\psfig{figure=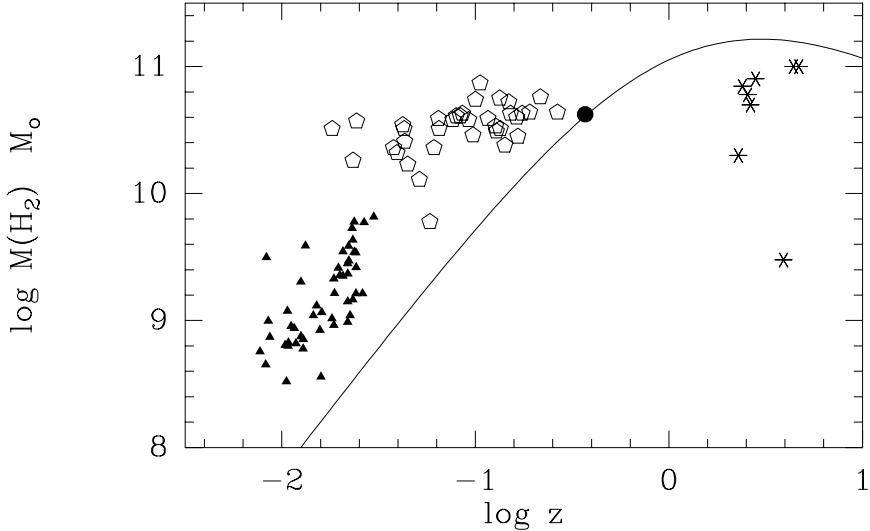,bbllx=2cm,bblly=1cm,bburx=12cm,bbury=16cm,width=12cm,angle=-90}
\caption{H$_2$ masses for the CO-detected objects at high redshift (filled
stars), compared to the ultra-luminous-IR sample of Solomon et al (1997, 
open pentagons), and to the Coma supercluster sample from Casoli et 
al (1996, filled triangles).
There is no detected object between 0.3 and 2.2 in redshift, except the quasar 
3c48, marked as a filled dot (Scoville et al 1993, Wink et al 1997). The curve
indicates the 3$\sigma$ detection limit of I(CO) = 1 K km/s at the IRAM-30m 
telescope (equivalent to an rms of 1mK, with an assumed $\Delta V$ = 300km/s).
The points at high $z$ can be detected well below this limit, since they are 
gravitationally amplified. }
\end{figure}

\section{Dust emission in star-forming galaxies}

Most of the previous sources, detected in the CO lines, had previously 
been detected in the dust continuum. At high redshift, it becomes easier to 
detect the dust emission, because of the large K-correction (e.g. Blain \& 
Longair 1993): the emission is roughly
varying as $\nu^4$ with the frequency $\nu$ in the millimeter range,
until the maximum around 60$\mu$m. At one mm, it is even easier to 
detect objects at $z = 5$ than $z = 1$. This has motivated deep searches 
in blank fields with sensitive instruments, since they should be dominated 
by high redshift objects, if they exist in sufficient numbers. 
Their detection will give information about the star-formation rate as
a function of redshift, a debated question: the maximum of
star-formation rate, found around $z $ =2 from optical studies (Madau et al 
1996) could shift to higher $z$ if dust is obscuring the higher-redshift
objects. From recent infrared lines observations (Pettini et al 1998),
it does not seem a serious problem, however.

The first search was made with the SCUBA bolometer
on JCMT (Hawaii) towards  a cluster of galaxies, thought to serve as a 
gravitational lens for high-$z$ galaxies behind (Smail et al 1997). The 
amplification is in average a factor 2. A large number of sources were found, 
all at large redshifts ($z > 1$), extrapolated 
to 2000 sources per square degree (above 4mJy), 
revealing a large positive evolution with redshift, i.e. an increase of 
starbursting galaxies. Searches toward the Hubble Deep Field-North
(Hughes et al 1998), and towards the Lockman hole and SSA13 (Barger et al 1998),
have also found a few sources, allowing to derive a similar density of sources:
800 per square degree, above 3 mJy at 850 $\mu$m. This already can account for 
50\% of the cosmic infra-red background (CIRB), that has been estimated by
Puget et al (1996) and Hauser et al (1998) from COBE data.
The photometric redshifts of these sources range between 1 and 3.
Their identification with optical objects might be  uncertain (Richards 1998).
However, Hughes et al (1998) claim that the star formation rate derived from
the far-infrared might be in some cases 10 times higher than derived
from the optical, due to the high extinction. If  only some of the sources
have a redshift higher than 4, it will flatten the Madau curve at high $z$.

Eales et al (1999) surveyed some of the CFRS fields at 850$\mu$m
 with SCUBA and found also that the sources can account for a significant
fraction of the CIRB background ($\sim$ 30\%). Their interpretation in terms 
of the star formation history is however slightly different, in that they
do not exclude that the submm luminosity density could evolve in the same
way as the UV one.
Deep galaxy surveys at 7 and 15$\mu$m with ISOCAM also see an evolution with 
redshift of star-forming galaxies: heavily extincted starbursts represent less
than 1\% of all galaxies, but 18\% of the star formation rate out to $z = 1$ 
(Flores et al 1999).

\section{Molecules in absorption}

Absorption techniques are also very efficient in the millimeter range,
and a few systems have been discovered at high redshift, between
$z$ = 0.2 to 1 in the last years (Wiklind \& Combes 1994, 95, 96;
Combes \& Wiklind 1996). The 
sensitivity is such that a molecular cloud on the line of sight
of only a few solar masses is enough to detect a signal,
while in emission, upper limits at the same distance are of the order 
of 10$^{10}$ M$_\odot$. Some general properties of the known
absorbing systems are summarised in Table 2.
They reveal to be the continuation
at high column densities (10$^{21}$--10$^{24}$ cm$^{-2}$)
of the whole spectrum of absorption systems, from the
Ly$\alpha$ forest (10$^{12}$--10$^{19}$ cm$^{-2}$)
to the damped Ly$\alpha$ and HI 21cm absorptions
(10$^{19}$--10$^{21}$ cm$^{-2}$).

\begin{table}[h]
\caption{Properties of molecular absorption line systems at high $z$} 
\begin{center} 
\begin{tabular}{lcccccrc}
Source & z$_{\rm a}^{a}$ & z$_{\rm e}^{b}$ &
$N_{\rm CO}$ & $N_{\rm H_2}$ & $N_{\rm HI}$ &
A$_{V}^{\prime c}$ & $N_{\rm HI}/N_{H_2}$ \\
 & & & cm$^{-2}$ & cm$^{-2}$ & cm$^{-2}$ & & \\
\hline \\
PKS1413+357   & 0.24671 & 0.247  & $2.3 \times 10^{16}$ & $4.6 \times 10^{20}$
& $1.3 \times 10^{21}$ & 2.0 & 2.8 \\
B3\,1504+377A & 0.67335 & 0.673  & $6.0 \times 10^{16}$ & $1.2 \times 10^{21}$
& $2.4 \times 10^{21}$ & 5.0 & 2.0 \\
B3\,1504+377B & 0.67150 & 0.673   & $2.6 \times 10^{16}$ & $5.2 \times 10^{20}$
& $<7 \times 10^{20}$ & $<$2 & $<$1.4 \\
B\,0218+357   & 0.68466 & 0.94   & $2.0 \times 10^{19}$ & $4.0 \times 10^{23}$
& $4.0 \times 10^{20}$ & 850 & $1 \times 10^{-3}$ \\
PKS1830--211A & 0.88582 & 2.51      & $2.0 \times 10^{18}$ & $4.0 \times 10^{22}$
& $5.0 \times 10^{20}$ & 100 & $1 \times 10^{-2}$ \\
PKS1830--211B & 0.88489 & 2.51      & $1.0 \times 10^{16 d}$ &
$2.0 \times 10^{20}$ & $1.0 \times 10^{21}$ & 1.8 & 5.0 \\
PKS1830--211C & 0.19267 & 2.51      & $<6 \times 10^{15}$                   &
$<1 \times 10^{20}$ & $2.5 \times 10^{20}$ & $<$0.2 & $>$2.5 \\
\hline
\end{tabular}
\end{center}
$^{a}${Redshift of absorption line}
$^{b}${Redshift of background source}
$^{c}${Extinction corrected for redshift using a Galactic extinction law}
$^{d}${Estimated from the HCO$^{+}$ column density
of $1.3 \times 10^{13}$\,cm$^{-2}$}
$^{e}${21cm HI data taken from Carilli et al. 1992,
1993, 1998}
\end{table}

About 15 different molecules have been detected
in absorption at high redshifts, in a total of 30
different transitions. This allows a detailed
chemical study and comparison with local clouds
(Wiklind \& Combes 1997, Carilli et al 1998).
Up to now, no significant variations in abundances
have been found as a function of redshift, at
least within the large intrinsic dispersion
already existing within a given galaxy.
Note that the high redshift allows us to
detect some new molecular lines, never observed from the ground
at $z = 0$, such as O$_2$ (Combes et al 1997), H$_2$O or LiH (Combes
\& Wiklind 1997, 1998). Molecular oxygen has not yet been detected
in space, and water vapour appears to be extended and cold (see also
the SWAS satellite preliminary results, Melnick et al 1999).

\subsection{ The LiH molecule}

Primordial molecules are thought to play a fundamental role in the 
early Universe, when stellar nucleosynthesis has not yet enriched 
the interstellar medium. After the decoupling
of matter and radiation, the molecular radiative processes, and 
the formation of H$_2$, HD and LiH contribute significantly 
to the thermal evolution of the medium (e.g. Puy et al 1993, Haiman,
Rees \& Loeb 1996). LiH has the lowest rotational first level
($\approx 21\,\rm K$ above the ground level)
and plays a unique role in the cooling of primordial clouds.
Unfortunately, the first transition of HD is
at very high frequency (2.7 THz), and the first LiH line, although
only at 444 GHz, is not accessible from the ground at $z=0$ due
to H$_2$O atmospheric absorption. 

Although the Li abundance is low (10$^{-10}$-10$^{-9}$), the observation
of the LiH molecule in the mm range is facilitated
by its large dipole moment, $\mu = 5.9$ Debye.
 It has been proposed that the LiH molecules could smooth the primary
CBR (Cosmic Background Radiation)
anisotropies, due to resonant scattering, or create secondary
anisotropies, and they could be the best way to detect primordial
clouds as they turn-around from expansion (e.g. Maoli et al 1996).
This motivated a search for
LiH at very high redshifts ($z \sim 200$), which 
resulted in upper limits (de Bernardis et al 1993).

The predictions of the LiH abundance changed drastically
in recent years, from the value of LiH/H$_2$ as high as 10$^{-6.5}$
corresponding to a primordial LiH/H ratio of $\sim$10$^{-12.5}$
(Lepp \& Shull 1984), to the value of LiH/H of $< 10^{-15}$ in 
the postrecombination epoch, from Stancil et al. (1996).
Indeed, quantum mechanical computations now predict
the rate coefficient for LiH formation through radiative association 
to be 3 orders of magnitude smaller than previously thought
from semi-classical methods (Dalgarno et al 1996).
In more evolved dense clouds, when three-body association reactions 
and formation of dust grains are
taken into account, a significant fraction of all lithium could
turn into molecules. 
 It is therefore important to try to detect LiH in a dense
cloud, at high enough redshift that the line falls in an
atmospheric window. This is the case for the absorbing cloud
in the lensing galaxy in front of the B0218+357 quasar 
at a redshift of $z=0.68466$. The H$_2$ column
density is estimated to be N(H$_2$)$ = 5 \times 10^{23}$ cm$^{-2}$
(Combes \& Wiklind 1997). From a tentative detection (see figure 2)
it was possible to deduce LiH/Li $\sim$ 1.5 10$^{-3}$. 
The uncertainty associated with the derived abundances are
large, but the low LiH/Li ratio seems to exclude complete transformation
of Li into LiH, as would be
expected in very dense clouds (e.g. Stancil et al 1996,
although the Li chemistry is not yet completely understood
in dark clouds).
However, it is likely that the cloud is clumpy, and in some of the more
diffuse parts, LiH is photodissociated;
some regions of the cloud could have a
higher excitation temperature, in which case our computation
under-estimates the LiH abundance.

\begin{figure}
\centering
\psfig{figure=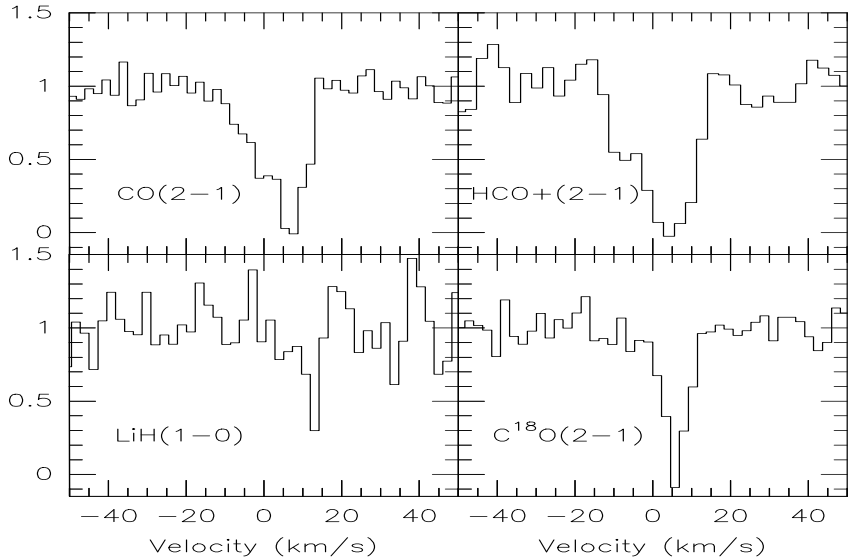,height=8cm,bbllx=3cm,bblly=6cm,bburx=16cm,bbury=20cm,width=12cm,angle=0}
\caption{ Tentative LiH detection towards B0218+357, at a redshift of
$z = 0.68466$. The line is compared to optically thick lines
like HCO$^+$(2-1) and CO(2-1), and more optically thin, such as 
C$^{18}$O(2-1), normalised to the absorbed continuum. }
\end{figure}

\subsection { Determination of the Hubble constant }

The best candidates to determine the Hubble
constant via gravitational lenses, are quasars for which the multiple
images are close together; then the lens can be only a bulge of a 
spiral galaxy, quite simple to model (contrary to widely separated images,
where the lens is a cluster of galaxies). But then, the light rays pass
close to the center of the lens galaxy, and absorption is likely to
occur. Among the four absorption systems detected in the mm
(cf Table 2), there are two cases of
confirmed gravitational lenses, with two images
(the two other cases are likely to be internal absorption).
For these systems (B0218+357 and
PKS1830-210), the separation between the two images is among
the smallest recorded: 0.3 and 1 arcsec respectively.
This makes these two objects good candidates for the
determination of the Hubble constant, and the monitoring
of the mm absorption can help to estimate the time-delay (Wiklind \& Combes
1995, 1998). Already, Lovell et al (1998) determined a time
delay of 26$^{+4}_{-5}$ days in PKS1830-210 from radio-cm observations.
Given the recently determined redshift of the quasar, of $z$ = 2.507
(Lidman et al 1999, in prep), the derived Hubble constant is 
65$^{+16}_{-9}$ km/s/Mpc. In B0218+357, Biggs et al (1999, in press)
were able to determine H$_0$ = 69$^{+7}_{-10}$ km/s/Mpc.

\subsection{Determination of CMB temperature}

Molecular lines in the millimeter domain are a tool to 
determine the background temperature, not because molecular clouds
are usually very cold, with a kinetic temperature
of the order of 10-20K, but mainly because the lines
are excited by collisions with H$_2$. In diffuse
regions, the collisional excitation is not enough, and
the radiative excitation dominates. The excitation temperature of the
molecules is then lower than the kinetic temperature,
and close or equal to the background temperature
$T_{bg}$. 
This is precisely the case of the gas absorbed in front of PKS1830-211, where
$T_{ex} \sim T_{bg}$ for most of the molecules. The measurement of
$T_{ex}$ requires the detection of two nearby transitions. When the
lower ones is optically thick, only an upper limit can be derived
for $T_{ex}$. Ideally, the two transitions should be optically thin,
but  then the higher one is very weak, and long integration times are
required.

The results obtained with the SEST-15m and IRAM-30m on different 
molecules agree and 
are plotted as a single point in
figure 3. Surprisingly, the bulk of measurements
points towards an excitation temperature lower than the background
temperature at $z=0.88582$, i.e. $T_{bg}$= 5.20K. 
We suspect that this could be due to a variation of the total 
continuum flux, due to a micro-lensing event.

\begin{figure}
\centering
\psfig{figure=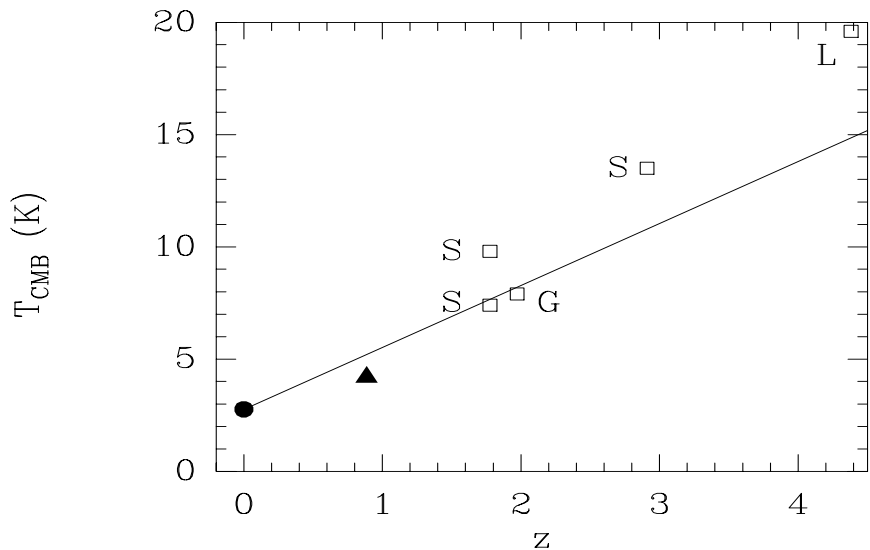,height=8cm,bbllx=2cm,bblly=1cm,bburx=12cm,bbury=14cm,width=12cm,angle=-90}
\caption{Summary of CMB temperature measurements as a function
of redshift. The filled dot is from COBE (Mather et al 1994). The squares are
upper limits obtained on the CI or CII from Songaila et al (1994ab, S), Lu et 
al (1996, L) and Ge et al (1997, G). Our point is the filled triangle. The 
line is the (1+z) expected variation. }
\end{figure}

\section{Predictions at even higher redshifts}

At the present time, only objects at $z$ below 10 have been detected.
The amount of star formation at higher redshifts will be negligible for
the global balance of the Universe, 
since the corresponding time is less than 3\% of the Hubble time; 
however, it is of primordial importance to trace back the first stars
formed, to understand star/galaxy formation processes just after 
recombination. Would it be possible, with the future millimetric
instruments, that will gain at least a factor 20 in sensitivity with respect
to the present ones, to detect
starbursts at even higher redshifts?

First, it will be easy to detect starbursts at $z \sim$ 5 without the
help of gravitational lensing as today (see figure 1). Beyond,
since we do not know the nature of the objects, we can extrapolate
their physical characteristics from the ultra-luminous FIR sample
detected at lower $z$. As can be seen in Table 1, molecular masses 
range from 10$^{10}$ to 10$^{11}$ M$_\odot$, the dust temperature
is high, about 30-50K (up to 100K), and their size is strikingly small,
below one kpc (300pc disks, Solomon et al 1997). In these conditions, 
the average column density of H$_2$ is 10$^{24}$ cm$^{-2}$, and the dust 
becomes optically thick at $\lambda <$ 150$\mu$m.  Two extreme simple 
models can be made about the corresponding molecular medium: either the gas 
is distributed in an homogeneous sphere, at a temperature of 50K, with a 
density of  10$^{3}$ cm$^{-3}$ in average, or the gas is clumpy, distributed 
in cold dense clouds, with embedded hot cores. The expected flux coming from 
such starbursting regions in the various CO lines, and the continuum flux 
from the dust emission, can then be derived as a function of redshift 
(cf Combes et al 1999). The results of the two-component model,
with kinetic temperatures 30 and 90 K at $z$ = 0, and total gas mass
6 10$^{10}$ M$_\odot$ is plotted in figure 4.

\begin{figure}[h]
\centering
\psfig{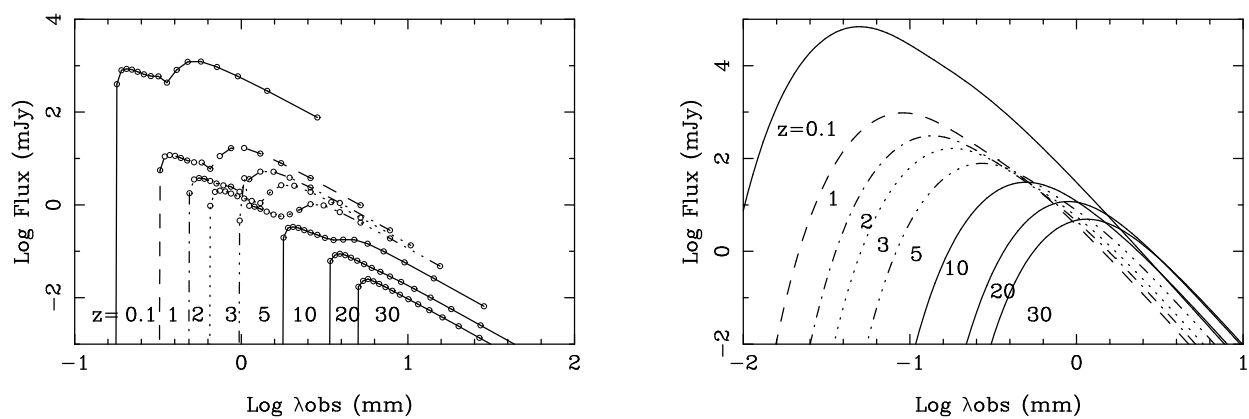}
\caption{ Expected flux for a starburst model in two-component clouds, 
of densities 10$^4$ and 10$^6$ cm$^{-3}$ for various
redshifts $z$ = 0.1, 1, 2, 3, 5, 10, 20, 30, and $q_0$ = 0.5.
Left are the CO lines, materialised
each by a circle. Right is the continuum emission from dust.}
\end{figure}

This picture shows that, since the low-$J$ CO lines are optically thick, 
their flux varies with the wavelength as $\lambda^{-2}$ at large
$\lambda$ in the millimeter range. This means that the CO-lines
detection will be less favoured at high 
redshift than the continuum emission. 
Unfortunately, the increase of the CMB temperature T$_{bg}$ with
redshift has not a significant effect on the detectability at
high $z$: if the excitation temperature of the gas is indeed
increased, the background emission to subtract is also increased
in the same proportions.
It is already much more difficult to detect 
objects at $z = 5$ than at $z = 1$, contrary to the dust emission.
This conclusion is opposite to what was found by Silk \& Spaans (1997)
in their model.
The most favourable wavelengths to detect the CO lines are always
longer than 1mm (assuming that the kinetic temperature of the gas is the
same as the dust temperature).
 We can note that it is already possible to detect now in the continuum such 
huge starbursts at any redshifts (up to 30). With the upcoming future 
millimeter instruments, it will be possible to detect them also in the
CO lines, or detect much more modest starbursts in the continuum.

%
  

\begin{references}  
\bibitem{}Barger A.J., Cowie L.L., Sanders D.B. et al.: 1998, Nature 394, 248
\bibitem{}Barvainis R., Alloin D., Guilloteau S., Antonucci R. 1998, ApJ 492, L13 
\bibitem{}Barvainis R., Tacconi L., Antonucci R., Coleman P.: 1994, Nature 371, 586
\bibitem{}Blain A.W., Longair M.S.: 1993, MNRAS 264, 509
\bibitem{}Brown R., Vanden Bout P.: 1992, ApJ 397, L19
\bibitem{}Brown R., Vanden Bout P.: 1991, AJ 102, 1956
\bibitem{}Carilli C.L., Perlman E.S., Stocke J.T.: 1992, ApJ 400, L13
\bibitem{}Carilli, C.L., Rupen, M.P., Yanny, B. 1993, ApJ 412, L59 
\bibitem{}Carilli, C.L., Menten K.M., Reid M.J., Rupen, M.P., Yun M.S.: 1998, ApJ 494, 175
\bibitem{}Carilli, C.L., Menten K.M., Reid M.J., Rupen, M.P., 
Claussen M.: 1998, in "Structure and Evolution of the Intergalactic Medium
from QSO Absorption Line Systems", ed. P. Petitjean and S. Charlot, Editions
Fronti\`eres, p. 325
\bibitem{}Casoli F., Dickey J., Kazes I. et al.: 1996, A\&AS 116, 193
\bibitem{}Combes F., Wiklind T., Nakai N.: 1997, A\&A 327, L17 
\bibitem{}Combes F., Wiklind T.: 1996, in "Cold gas at high redshift", ed. Bremer M., Rottgering H., van der Werf P., Carilli C.L. (Dordrecht:Kluwer), p. 215         
\bibitem{}Combes F., Wiklind T.: 1997, ApJ 486, L79 
\bibitem{}Combes F., Wiklind T.: 1998, A\&A 334, L81  
\bibitem{}Combes F., Maoli R., Omont M.: 1999, A\&A in press
\bibitem{}Dalgarno A., Kirby K., Stancil P.C.: 1996, ApJ 458, 397
\bibitem{}de Bernardis P., Dubrovich V., Encrenaz P. et al.: 1993, A\&A 269, 1
\bibitem{}Downes D., Neri R., Wiklind T., Wilner D.J., Shaver P.: 1998, ApJ preprint (astro-ph/9810111)
\bibitem{}Downes D., Solomon P.M., Radford S.J.E. 1995, ApJ 453, L65 
\bibitem{}Eales S.A., Lilly S.J., Gear W.K., Dunne L., Bond J.R., Hammer F., Le Fevre O., 
Crampton D.: 1999, ApJL in press (astro-ph/9808040)
\bibitem{}Flores H., Hammer F., Thuan X. et al. : 1999, ApJ in press (astro-ph/9811202)
\bibitem{}Frayer D.T., Ivison R.J., Scoville N.Z., et al., 1998, ApJ 506, L7
\bibitem{}Ge J., Bechtold J., Black J.: 1997, ApJ 474, 67
\bibitem{}Guilloteau S., Omont A., McMahon R.G., Cox P., PetitJean P.: 1997, A\&A 328, L1
\bibitem{}Haiman Z., Rees M.J., Loeb A.:1996, ApJ 467, 522
\bibitem{}Hauser M.G., Arendt R.G., Kelsall T., et al.: 1998, ApJ 508, 25
\bibitem{}Hughes D.H., Serjeant S., Dunlop J. et al.: 1998, Nature 394, 241
\bibitem{}Lepp S., Shull J.M.: 1984, ApJ 280, 465
\bibitem{}Lo K.Y., Chen H-W., Ho P.T.P.: 1999, A\&A 341, 348
\bibitem{}Lovell J.E., Jauncey D.L., Reynolds J.E. et al.: 1998, ApJ 508, L51
\bibitem{}Lu, L., Sargent, W. L. W., Womble, D. S.,  Barlow, T. A.: 1996, ApJ, 457, L1
\bibitem{}Madau P., Ferguson H.C., Dickinson M.E. et al.: 1996, MNRAS 283, 1388
\bibitem{}Maoli R., Ferruci V., Melchiorri F. et al: 1996, ApJ 457, 1
\bibitem{}Mather, J. C., et al.: 1994, ApJ, 420, 439
\bibitem{}Melnick G.J. et al.: 1999, BAAS 193, 7201
\bibitem{}Ohta K., Yamada T., Nakanishi K., Kohno K., Akiyama M., Kawabe R.: 1996, Nature 382, 426
\bibitem{}Omont A., Petitjean P., Guilloteau S., McMahon R.G., Solomon P.M.: 1996, Nature 382, 428
\bibitem{}Pettini M., Kellog M., Steidel C.C. et al.: 1998, ApJ 508, 539
\bibitem{}Puget J.L., Abergel A., Bernard J-P. et al. : 1996, A\&A 308, L5
\bibitem{}Puy D., Alecian G., Le Bourlot J., L\'eorat J., Pineau des
For\^ets G., 1993, A\&A 267, 337
\bibitem{}Richards E.R.: 1998, ApJ in press (astro-ph/9811098)
\bibitem{}Scoville N.Z., Padin S., Sanders D>B> et al. : 1993, ApJ 415, L75
\bibitem{}Scoville N.Z., Yun M.S., Windhorst R.A., Keel W.C., Armus L.: 1997, ApJ 485, L21
\bibitem{}Silk J., Spaans M.: 1997, ApJ 488, L79
\bibitem{}Smail I., Ivison R.J., Blain A.W.: 1997, ApJL 490, L5
\bibitem{}Solomon P.M., Downes D., Radford S.J.E., Barrett J.W.: 1997, ApJ 478, 144
\bibitem{}Solomon P.M., Downes D., Radford S.J.E.: 1992, Nature 356, 318
\bibitem{}Songaila, A., Cowie L.L., Hogan C., Rugers M.: 1994a:  Nature, 368, 599
\bibitem{}Songaila, A., Cowie L.L., Vogt S. et al.: 1994b: Nature, 371, 43
\bibitem{}Stancil P.C., Lepp S., Dalgarno A.: 1996, ApJ 458, 401
\bibitem{}Yun M.S., Scoville N.Z.: 1998, ApJ 507, 774
\bibitem{}Wiklind T., Combes F.: 1994, A\&A 286, L9 
\bibitem{}Wiklind T., Combes F.: 1995, A\&A 299, 382 
\bibitem{}Wiklind T., Combes F.: 1996, Nature 379, 139  
\bibitem{}Wiklind T., Combes F.: 1997, A\&A 328, 48 
\bibitem{}Wiklind T., Combes F.: 1998, ApJ 500, 129 
\bibitem{}Wink J.E., Guilloteau S., Wilson T.L.; 1997, A\&A 322, 427
\end{references}
\end{document}